\documentclass[aps,amsfonts,amsmath,prd,preprint,nofootinbib]{revtex4}
\newcommand{\beq}{\begin{equation}}
\newcommand{\eeq}{\end{equation}}
\usepackage{graphicx}
\usepackage{dcolumn}
\usepackage{bm}
\usepackage[colorlinks=true]{hyperref}
\usepackage{color}
\usepackage{multirow}
\begin{document}

\title{Cosmic strings and primordial black holes}


\author{Alexander Vilenkin$^a$ , Yuri Levin$^{b,c, d}$, and Andrei Gruzinov$^e$ }

\affiliation{$^a$ Institute of Cosmology, Department of Physics and Astronomy,
Tufts University, Medford, MA 02155}

\affiliation{$^b$ Center for Theoretical Physics, Department of Physics, Columbia University, New York, NY 10027}
\affiliation{$^c$ Center for Computational Astrophysics, Flatiron Institute, New York, NY 10010}
\affiliation{$^d$ School of Physics and Astronomy, Monash University, VIC 3800, Australia}

\affiliation{$^e$ Center for Cosmology and Particle Physics,  Department of Physics, New York University, New York, NY 10001}

\vskip .3in

\begin{abstract}
Cosmic strings and primordial black holes (PBHs) commonly and naturally  form in many  scenarios describing the early universe. Here we show that if both
 cosmic strings and PBHs are present, their interaction leads to a range of interesting consequences.
 At the time of their formation, the PBHs get attached to the strings and influence their evolution, leading to the formation
 of black-hole-string networks and commonly to the suppression of loop production in a
 range of redshifts. Subsequently, reconnections within the network give rise to small nets made of several black holes and connecting  strings.
 The number of black holes in the network as well as the stability of the nets depend on the topological properties of the strings.
 The nets oscillate and shrink exponentially due to the emission of gravitational waves.
This leads to potentially observable string-driven mergers of PBHs.
The strings can keep PBHs from galactic halos, making the current bounds on PBHs not generally applicable. Alternatively, heavy PBHs can drag low-tension
strings into the centers of galaxies. The superconducting strings can appear as 
radio filaments pointing towards supermassive black holes.
\end{abstract}
\maketitle

\section{Introduction}

Primordial black holes (PBHs) could be formed early in the radiation era and could have played an important role in the evolution of the universe (see Refs.~\cite{Carr:2009jm,Sasaki:2018dmp} for recent reviews).  They could serve as seeds for supermassive black holes observed at the centers of most galaxies and could account for merging BH binaries currently observed by LIGO.  It has also been suggested that PBHs could constitute a substantial part, if not all, of dark matter.

A seemingly unrelated direction of cosmological research is the formation and evolution of topological defects, in particular cosmic strings.  Strings are predicted in a wide class of particle physics models and can produce a variety of astrophysical effects (see \cite{book,Copeland} for a review).  Some models predict more complicated defects -- monopoles connecting by strings, with $N\geq 2$ strings attached to each monopole.  In the case of $N=2$ the defects are called "necklaces", with monopoles and antimonopoles alternating like beads along the string.  A discovery of cosmic strings would open a window into physics of very high energies, and new observational bounds on strings may rule out classes of particle physics models.

In this paper we discuss what happens when both strings and PBHs are present in the universe.  We find that they generically form a network where PBHs are connected by strings.  The evolution and observational effects of strings may then be significantly affected by PBHs and vice versa.  In particular, observational bounds on the string energy scale are strongly enhanced in some scenarios.  Another interesting effect is the increased merger rate for PBHs
connected by strings.


\section{Black hole formation and string capture}

\subsection{PBH formation}

PBHs can be formed by a variety of mechanisms.  The most widely discussed scenario is the collapse of primordial overdensities during the radiation era \cite{Zeldovich,CarrHawking}.  The initial overdensities could originate from quantum fluctuations in the inflationary epoch.  The Jeans mass at time $t$ during the radiation era is $M_J\sim t/G$, so black holes of mass $M$ form at 
\beq
t_f\sim GM \sim 10^{-5}\frac{M}{M_\odot} s, 
\eeq
with their Schwarzschild radius comparable to the cosmological horizon.  

The density fluctuation required for a horizon-size region to collapse to a black hole is $\delta\equiv \delta\rho/\rho\sim 1$.  On the other hand, the rms fluctuation on scales accessible to CMB and large-scale structure observations is $\delta_{rms}\sim 10^{-5}$, and the probability of having $\delta\sim 1$ on such scales is negligibly small.  Hence one has to assume that the fluctuation amplitude is strongly enhanced in the range of scales corresponding to PBH formation.  Here we shall assume for simplicity that the fluctuation spectrum has a narrow peak, so that all PBHs form at about the same time and have nearly the same mass.  Such fluctuation spectra naturally arise in some inflationary models (e.g., \cite{GarciaBellido:1996qt,Yokoyama:1995ex,Frampton:2010sw}).  

Suppose the fraction of horizon regions that turn into BHs at time $t_f$ is $\lambda\ll 1$.  Then the BH density at formation is $n_f\sim \lambda t_f^{-3}$ and their average separation is $d_f\sim\lambda^{-1/3}t_f$. 
The cold dark matter (CDM) mass density at that time is
\beq
\rho_{CDM}\sim\frac{1}{Gt_f^2} \left(\frac{t_f}{t_{eq}}\right)^{1/2},
\eeq
so the fraction of dark matter in the form of BHs is
\beq
f_M\sim \frac{Mn_f}{\rho_{CDM}}\sim \lambda\left(\frac{M_{eq}}{M}\right)^{1/2},
\label{flambda}
\eeq
where $M_{eq}\sim t_{eq}/G\sim 10^{17}M_\odot$ is the horizon mass at the time $t_{eq}$ of equal matter and radiation densities.  The quantity $f_M$ does not change with time and is therefore a useful characteristic of PBHs.

Let us now indicate some cosmologically interesting values of the parameters $M$ and $\lambda$.  There are only two observationally allowed windows for the mass $M$ where PBHs may account for all dark matter \cite{Carr:2016drx,Inomata:2017vxo}: $M\sim 10^{-15}-10^{-10} M_\odot$ and $M\sim 10-100 M_\odot$.  With $f_M\sim 1$, Eq.~(\ref{flambda}) then gives $\lambda\sim 10^{-13}-10^{-16}$ and $\lambda\sim 10^{-8}$, respectively.  In order to account for LIGO observations, we need $M\sim 10 M_\odot$ and \cite{Sasaki:2016jop} $f_M\sim 10^{-3}$, which gives $\lambda\sim 10^{-11}$.  For PBHs to serve as seeds of supermassive BHs, we need $M\sim 10^3-10^6 M_\odot$ and the comoving PBH density $\sim 0.1 Mpc^{-3}$.  The corresponding range of $\lambda$ is $\lambda\sim 10^{-12}-10^{-16}$. 

We now mention some other mechanisms of PBH formation.  High-energy vacuum bubbles may nucleate and expand during inflation, resulting in a very wide spectrum of bubble sizes.  After inflation ends, the bubbles collapse to form BHs \cite{GVZ,DV}.  Small bubbles collapse to BHs much smaller than the horizon, but starting with a certain critical size the BHs form with their Schwarzschild radius comparable to the horizon.  A closely related scenario is PBH formation by collapse of vacuum domain walls \cite{Garriga:1992nm,Khlopov:2004sc,Yoo,GVZ,DGV}.  Once again, sufficiently large walls form BHs at the horizon scale.  PBHs could also be formed by collapse of cosmic string loops \cite{Hawking:1987bn,Polnarev:1988dh,Garriga:1992nm}, but this requires the loops to be nearly circular and the PBH density produced in this way is not cosmologically interesting.

PBH formation can be enhanced if the universe went through some episodes of reduced pressure.  This could happen at cosmological phase transitions or during a period of slow reheating after inflation (see, e.g., \cite{Carr:2017edp} and references therein).  In this case the Jeans mass may be strongly reduced and the typical black hole size may be significantly smaller than the horizon.  This mechanism, however, is likely to operate only in the very early universe\footnote{We do not expect any phase transitions later than the QCD  transition at $t\sim 10^{-4}$~s.} and may only account for PBHs with $M\lesssim M_\odot$.

In what follows we shall assume that PBHs are formed on the horizon scale.  We shall comment on the alternative possibility at the end of the next subsection.

\subsection{String evolution and capture}

Cosmic strings are characterized by the energy scale of symmetry breaking, $\eta\ll m_p$, where $m_p \sim 10^{19}$~GeV is the Planck mass.  The mass per unit length of string is $\mu\sim \eta^2$ and the string tension is $T=\mu$.  The strings carry a magnetic flux of a heavy gauge field (which acquires a mass $\sim \eta$ due to the symmetry breaking).  When two strings cross, they reconnect in such a way that the direction of the magnetic flux is preserved along the strings.
The strength of string gravity is set by the dimensionless parameter $G\mu\sim (\eta/m_p)^2 \ll 1$.  
The strings are formed at a symmetry breaking phase transition at $t_s \sim (G\mu)^{-1} t_p$, where $t_p\sim 10^{-44}$~s is the Planck time.  Note that this is earlier than PBH formation for all astrophysically interesting values of $\eta$ and $M$.  

Numerical simulations of string evolution indicate that strings evolve in a scale-invariant manner \cite{Bennett,Allen,BlancoPillado:2011dq}.  A Hubble-size volume at any time $t$ contains a few long strings stretching across the volume and a large number of closed loops of length $l\ll t$.  Long strings move at mildly relativistic speeds, $v\sim 0.2$, so they typically have a chance to intersect with other strings in a Hubble time.  Loops are chopped off the long strings with a typical size $l\sim 0.1 t$.  Much smaller loops are also produced in localized regions where the string velocity approaches the speed of light.
The loops oscillate periodically, lose their energy by gravitational radiation, and gradually shrink and disappear.  Long string reconnections and chopping off of loops are both necessary to sustain the scale-invariant nature of evolution.

The lifetime of a loop of initial length $l$ is  
\beq
\tau_l \sim \frac{l}{\Gamma G\mu},
\eeq
where $\Gamma\sim 50$ is a numerical coefficient.  Gravitational waves emitted by loops over the cosmic history add up to a stochastic gravitational wave background.   Requiring that the predicted amplitude of this background is not in conflict with the millisecond pulsar observations, one can impose an upper bound on the string mass parameter \cite{Blanco-Pillado:2017rnf}:
\beq
G\mu\lesssim 10^{-11}.
\eeq

Suppose now that PBHs are formed at some time $t_f \gg t_s$.  To see what effect this will have on strings, consider an overdense region with $\delta \sim 1$ which is destined to become a BH.  While this region is bigger than the horizon, its evolution is unremarkable: both the background radiation density and the strings evolve just as they would in a region of a somewhat denser homogeneous universe.  When the region comes within the horizon at $t\sim t_f$, it finds itself surrounded by a domain of significantly lower density.  At that point an apparent horizon forms, encompassing more or less the entire region, so the region's interior becomes a BH.  The size of the region at that time is comparable to the cosmological horizon, so it typically contains $\sim 10$ long strings.  These strings suddenly find themselves partially engulfed by a BH.  Thus, we expect $\sim 10$ long strings to be captured in each BH.  In other words, there will be $\sim 20$ long strings coming out of the BH, with half of them carrying flux in and the other half out of the BH.

Long strings have the shape of random walks of step $\sim t$.  At $t\sim t_f$, the probability for a string to encounter a BH at each step along its length is $\sim \lambda$.  Hence the average length of string connecting two BHs is ${l}\sim t_f/\lambda$.  In fact, there will be very few strings much longer than that (since the probability of not encountering a BH for $n\gg \lambda^{-1}$ steps is exponentially suppressed).  The average initial separation of BHs connected by a string is $\sim \lambda^{-1/2}t_f$.  The evolution of this BH-string network will be discussed in the next section.

We finally comment on the scenarios where BHs have size much smaller than the horizon at formation.  Strings will be captured in this case only when the center of a newly formed BH happens to be within a Schwarzschild radius from a string, and it is unlikely for a BH to be attached to more than two strings.  If $n_f$ is the BH density at formation and $M\ll t_f/G$ is their mass, the initial density of BHs attached to the network is then
\beq
n\sim\left(\frac{GM}{t_f}\right)^2 n_f.
\eeq
This is equivalent to the above scenario with the parameter $\lambda$ replaced by $(GM/t_f)^2 \lambda$.  Otherwise, the evolution of the network is similar to that with BHs formed at the horizon scale.

\section{Evolution of BH-string network}

\subsection{The minimal scenario}

Strings emanating from any given BH typically change their shape and direction on a Hubble time scale.  With $\sim 20$ strings attached to the BH, they will frequently intersect and reconnect.  When intersecting strings have opposite directions of the magnetic flux, such reconnections leave a string segment with both its ends attached to the BH, while the reconnected long string pieces move away.  The segment will oscillate, self-intersect and chop off some closed loops.  Eventually what remains of the segment will be swallowed by the BH.  This process reduces the number of strings attached to the BH by two.  Another possibility to consider is when a string moving through the horizon volume intersects two of the strings attached to the BH.  Once again, if the flux directions in the attached strings are opposite, this will result in the formation of a sub-horizon attached segment.  The rate of these processes will go down as the number of attached strings decreases, but still most BHs may lose all their strings in this way.  We shall argue, however, that at any time there is a nonzero density of BHs that remain connected to the string network.

At $t\sim t_f$ the BH separation is much larger than the horizon.  So the string dynamics is largely unaffected by BHs, except in the rare horizon regions that do contain BHs.  Let us first recall that a string network without BHs exhibits a scaling evolution, with $\sim 10$ strings of length $\sim t$ per volume $t^3$, so the energy density in long strings is
\beq
\rho_s(t)\sim 10\frac{\mu}{t^2}.
\label{Lt}
\eeq 
The total energy of long strings in a large comoving volume decreases as 
\beq
{\cal E}(t) \propto \rho_s(t){a^3(t)} \propto t^{-1/2},
\eeq
where $a(t)\propto t^{1/2}$ is the scale factor. The total string length decreases in the same way, with the remaining length going into closed loops.\footnote{There is no significant stretching of strings by the expansion during the radiation era.}  

At $t\sim t_f$, all BHs are attached to the string network.  At later times, only a fraction $f(t) \sim (t_f/t)^{1/2}$ of the network length still survives.  If we assume that the loops are chopped off completely at random and that this process is unaffected by the presence of BHs, then the fraction of BHs that are still attached to the network at time $t$ is also $\sim f(t)$.  We shall refer to this as "the minimal scenario".   (In the next subsection we shall argue that the fraction of BHs remaining in the network may actually be significantly larger.)

According to the minimal scenario, the number density of BHs in the network is
\beq
n(t)\propto f(t)a^{-3}(t)\propto t^{-2}
\eeq
and the distance between them is $d(t)\propto t^{2/3}$.  Hence we can write
\beq
d(t)\sim \lambda^{-1/3} t_f^{1/3}t^{2/3}
\label{dt}
\eeq

The average length of string $l(t)$ per BH in the network at time $t$ can be found from
\beq
\rho_s(t)\sim \mu n(t)l(t).
\label{Ln}
\eeq 
This shows that $l(t)$ does not change with time, $l(t)\sim t_f/\lambda$.  This length becomes comparable to the horizon at $t_h\sim t_f/\lambda$, and it follows from Eq.~(\ref{dt}) that the BH separation $d(t_h)$ at that time is also $\sim t_h$.  Thus, at $t\sim t_h$ the network consists of BHs separated by distances $\sim t_h$ and connected by more or less straight strings.  The subsequent evolution depends on the average BH velocity $v_h$ at time $t_h$.

BHs are pulled by the strings with a force $\sim \mu$.  The typical BH velocity at time $t$ is 
\beq
v\sim (\mu/M)t \sim G\mu \frac{t}{t_f}
\label{vt}
\eeq
or $v\sim 1$, whichever is smaller.  If $G\mu\ll \lambda$, then $v_h \sim G\mu/\lambda \ll 1$, the distance travelled by BHs in a Hubble time is $\ll t_h$, so BHs are nearly stationary.  Then at $t> t_h$ the strings connecting BHs become nearly straight, and at later times they are simply stretched by the expansion.  The BH separation in this regime is 
\beq
d(t)\sim (t_h t)^{1/2}.
\eeq

The distance travelled by a non-relativistic BH in a Hubble time is $\sim (\mu/M)t^2$.  It becomes $\sim d(t)$ at\footnote{We assume that $t_h, t_d < t_{eq}$.}
\beq
t_d \sim (G\mu)^{-2/3}\lambda^{-1/3} t_f.
\eeq
The BH velocity at that time is $v_d \sim (G\mu/\lambda)^{1/3}\ll 1$.  The BHs are still moving non-relativistically, but they can no longer be regarded as stationary.  
The strings will now cross and reconnect, chopping off loops and finite BH-string nets of size $\sim d(t)$.  For $t>t_d$, the BHs move a distance larger than their current separation in the network. This should lead to string intersections, as a result of which some BHs will disconnect, and the new BH separation {\it in the network} will be comparable to the distance traveled by the BHs: 
\beq
d(t)\sim v(t)t \sim \frac{\mu t^2}{M},
\label{dt2}
\eeq
where we have used Eq.~(\ref{vt}) for $v(t)$.  The network configuration in this regime is changing on the Hubble timescale $\sim t$.

The BHs start moving relativistically at $t\sim t_*$, where
\beq
t_* \sim \frac{M}{\mu} \sim 10^{-5} (G\mu)^{-1} \frac{M}{M_\odot} s .
\eeq
At this time the network reaches equipartition, $\rho_s\sim\rho_{BH}$.  At later times, the scale of the network is 
\beq
d(t)\sim t.
\eeq
However, the equipartition between strings and BHs would require the BHs move with increasingly large Lorentz factors.  Accelerating to ultra-relativistic velocities would require a large degree of coherence in the direction of the tension force acting on a BH, and does not generally take place. Instead, at $t>t_d$ the BHs become energetically subdominant compared to strings.
Note that the evolution of the infinite network at $t>t_d$ is independent of the initial BH density (which is characterized by the parameter $\lambda$).

If $G\mu >\lambda$, then $v_h\sim 1$, so BHs become relativistic right after $t_h$. 

The key assumption in the above scenario is that in the early stages of evolution the loops are chopped off at random, regardless of whether or not the loop contains a BH.  In the next section we shall argue that chopping off a loop with a BH may be considerably more difficult, so our estimates can be regarded as lower bounds on the number of BHs in the network.

\subsection{Black hole detachment}

Loops are chopped off the long strings by collision of wiggles traveling along the string in opposite directions.\footnote{Large loops of size $\sim t$ are occasionally formed by self-intersection of long Brownian strings.  Such loops either reconnect to the network or fragment into smaller loops.  Simulations indicate that this mechanism of loop formation is subdominant.}  Suppose now there is a BH attached to the string and let us assume for simplicity that there is only one other string attached to that BH.  We shall assume also that the BH is so massive that it is not moved much (in a Hubble time) by the forces exerted by the strings.  This is a reasonable assumption at early stages of the evolution. 

A loop that includes the BH can only be formed if the two strings emanating from the hole cross one another.
It can be shown that a wave traveling along a straight string towards a stationary BH is simply reflected by the BH (approximating the BH as a point where the string is pinned at a fixed position).  So if the two strings come out of the BH at a substantial angle from one another, waves traveling along the strings are not likely to collide.  However, a collision becomes very likely if the angle between the two strings is sufficiently small.

Simulations indicate that strings in the network consist of nearly straight segments separated by kinks \cite{BOS}.  The kinks travel along the strings at the speed of light.  Most kinks are rather mild, so the string direction varies very little from one segment to another.  But there are also some large kinks, which are separated by distances $\sim t$ along the strings.  Now consider a string attached to a BH.  As kinks travel along the string towards the BH, the direction of the string does not change much -- until a large kink arrives.  Such sudden, large change of direction will occur about once in a Hubble time.
Suppose the angle at which reconnection of two strings becomes likely is $\sim \pi/6$.  This corresponds to a solid angle $\sim 0.1$.  Then we expect the strings to be in a "dangerous" configuration about every 10 Hubble times.  Now, the number of Hubble times between some initial time $t_1$ and some later time $t \gg t_1$ is $N\sim \ln (t/t_1)$.  Hence, a BH attached to two strings at $t_1$ can be expected to disconnect at $t\sim e^{N} t_1 \sim 10^4 t_1$.  If $f(t)$ is the fraction of BHs remaining in the network at time $t$, then we must have $f(e^N t)\sim e^{-1} f(t)$.  This has the solution $f(t)\propto t^{-1/N}$, which decreases much slower than $f(t)\propto t^{-1/2}$ assumed in the minimal scenario.

This is of course only a rough estimate, but it does suggest that BH survival in the network may be more likely than the naive analysis in the preceding section indicates.  A reliable determination of the fraction of surviving BHs would require a numerical simulation of the string-BH network.
 
{We note also that strings of a special kind, called necklaces, are more likely to remain attached to BHs than "ordinary" strings.}
Consider a sequence of symmetry breaking phase transitions $G\to H\times U(1)\to H\times Z_2$, where $G$ is a semisimple group and $H$ is its subgroup.  Then monopoles carrying the magnetic flux of $U(1)$ are formed in the first phase transition, and their flux is squeezed into strings at the second phase transition, so that each monopole  gets attached to two strings.  (The magnetic charge of the monopole is twice the flux carried by the string.) The monopoles will then be like beads on the strings.  These defects are quite generic \cite{Kibble:2015twa}.  The evolution of necklaces has been studied in Refs.~\cite{necklace,BPO}, with the conclusion that relativistic string motion causes the monopoles to move along the strings at speeds close to the speed of light.  As a result, monopole-antimonopole annihilation on the string is rather efficient, and the separation between them on a long string is comparable to the horizon.  Monopoles therefore have little effect on the string dynamics and evolution.
When a BH is formed, it captures some monopoles and antimonopoles together with the strings.  With $N\sim 10$ strings within the Hubble volume, we can expect the total magnetic charge of the BH to be $2\sqrt{10}\sim 6$, so there should be at least $\sim 6$ strings attached to it.  These strings have the same direction of the magnetic flux, so their reconnections do not change the number of the attached strings.\footnote{``Dangerous" string crossings that can result in BH detachment are likely to occur close to the BH, e.g., within a distance $\sim 0.1 t$.  We can expect that the strings will be cleared of monopoles in this range, so the string orientation will not reverse.}  



\subsection{Persistent networks: the maximal scenario}

Considering the possibility that the minimal scenario may overestimate the ease at which BHs are detached from the network, we shall now discuss the opposite limiting case, assuming that a substantial fraction of BHs remain attached to the network.  We shall refer to this as the maximal scenario.

In this scenario, the BH number density in the network is
\beq
n(t)\sim \frac{\lambda}{(t_f t)^{3/2}},
\eeq
so the average BH separation is
\beq
d(t)\sim\lambda^{-1/3}(t_f t)^{1/2}.
\eeq
At early times, when $d(t)\gg t$, the average length of strings can be found from Eqs.~(\ref{Lt}),(\ref{Ln}):  
\beq
l(t)\sim \frac{1}{t^2 n(t)}.
\eeq
At $t_h\sim \lambda^{-2/3}t_f$ the BH separation becomes comparable to the horizon.  The average string length at that time is also $\sim t_h$.

The typical BH velocity at time $t_h$ is
\beq
v_h\sim (\mu/M)t_h \sim \lambda^{-2/3}G\mu.
\eeq
For most of the astrophysically interesting parameter values that we mentioned in Sec.~II we have $v_h\ll 1$.  Assuming this is the case, at $t>t_h$ the strings are stretched by the expansion and become nearly straight.
 
As long as BHs move non-relativistically, their displacement during a Hubble time $t$ is $\sim (\mu/M)t^2$.  It becomes $\sim d(t)$ at
\beq
t_d\sim\lambda^{-2/9}(G\mu)^{-2/3}t_f.
\eeq
The BH velocity at that time is 
\beq
v_d\sim (\mu/M)t_d\sim v_h^{1/3}.
\eeq 
The average kinetic energy of BHs at $t>t_d$ is $\sim \mu d(t)$.  It becomes $\sim M$ at $t\sim t_*$, when $\mu d\sim M$,
\beq
t_*\sim \lambda^{2/3}(G\mu)^{-2}t_f \sim v_d^{-4}t_d.
\eeq
At this point BHs start moving relativistically. 
The energy density of the BH-string network at $t>t_*$ is $\rho_{net}\sim \mu d^{-2}(t)$ and
\beq
\frac{\rho_{net}}{\rho_r}\sim \lambda^{2/3} G\mu\frac{t}{t_f},
\eeq
where $\rho_r \sim (Gt^2)^{-1}$ is the radiation density.  The network dominates the energy density of the universe at 
\beq
t_{dom}\sim \lambda^{-2/3}(G\mu)^{-1} t_f.
\eeq

There is a range of possible evolution scenarios between the minimal and maximal models.  For example, the network may evolve as in the maximal scenario until $t\sim t_d$, but at later times multiple string crossings in the vibrating BH-string network could result in copious production of closed loops and small nets.  Then the subsequent evolution could be similar to that in the minimal scenario.  Numerical simulations will be necessary to explore these possibilities.

\section{Black holes with loops, and  nets}

The spectrum of stochastic gravitational radiation produced by oscillating  loops of string is determined by the size distribution of loops.  In the presence of PBHs this distribution is rather different from that in the standard cosmic string scenario, where $\sim 10$ loops of length $\sim 0.1 t$ are produced per Hubble time $t$ in a Hubble volume $t^3$.
In a BH-string network, the loop production is initially the same as in the standard scenario, but then it ceases completely 
at $t\sim t_h$.  In the minimal model the loop production resumes at $t\sim t_d$, with about one loop of size $\sim d(t)$
formed per Hubble time in a volume $\sim d^3(t)$, where $d(t)$ is from Eq.~(\ref{dt2}).  Apart from closed string loops, the networks will yield other gravitational wave sources -- oscillating string segments attached to BHs and small BH-string nets.

At $t>t_d$ the BH size is much smaller than $d(t)$, so a string segment attached to a BH can be approximately described as a segment with both ends pinned at a fixed point in space.  A segment of invariant length $l$ would then oscillate with a period $T=2l$, assuming that it does not self-intersect.  We, together with Elena Murchikova,  have studied numerically a  number of such pinned segment configurations and found, somewhat surprisingly, 
that most of them self-intersect\footnote{This is in contrast to string loops without BHs, where simple families of loop solutions are non-intersecting in a substantial portion of the parameter space.}. 
We did find some examples of pinned loops that did not self-intersect, but it is far from clear whether these are generic. It is possible that a generic pinned loop, after several
self-intersections and pinching off the "looplets", will evolve into
a non-self-intersecting configuration. This would endow a PBH with an oscillating string ``hair". 

PBHs with attached string loops can be of astrophysical interest.  For example, if the PBHs were captured into halos of galaxies they would drag the loops attached to them into the Milky Way and therefore make the gravitational waves produced by the loops more detectable. 
An exploration of the  dynamics and gravitational-wave signatures of  loops attached to BHs will be published elsewhere.

\begin{figure}
\includegraphics[width=\columnwidth]{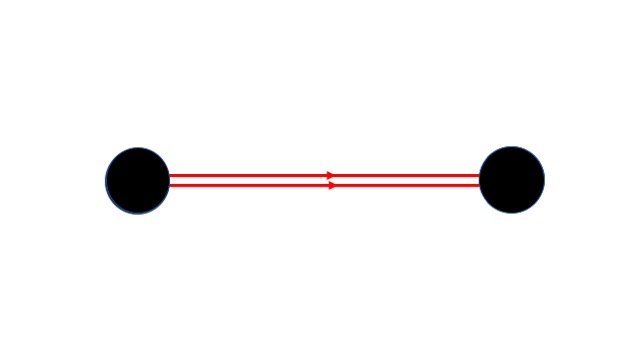}
    \caption{A stable binary: two black holes with charges $2$ and $-2$ are connected by $2$ strings. The connections are indestructible by the internal dynamics of the binary. The binary shrinks by emitting gravitational waves.}
    \label{fig:binary}
\end{figure}

Another interesting object is a small net consisting of a few BHs connected by strings.  The simplest net is a binary system consisting of two BHs connected by two strings, see Fig.~(\ref{fig:binary}).  In an oscillating binary the strings will frequently intersect and the binary will break up if the strings have opposite flux directions.  For ``ordinary" strings, the fluxes in the two strings should be opposite, but for necklaces the fluxes can be the same if the BHs have opposite magnetic charges.  
In the latter case the binary  is  long-lived.

\begin{figure}
\includegraphics[width=\columnwidth]{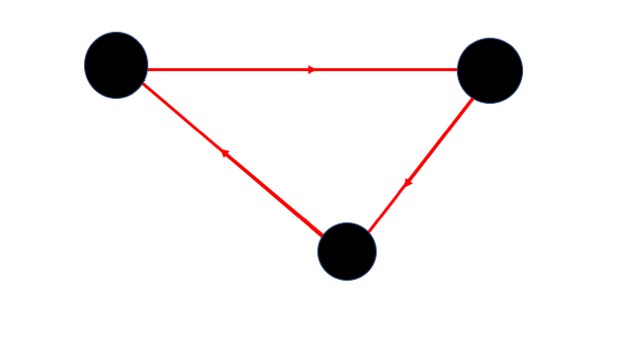}
    \caption{An oscillating triangle: three non-relativistically moving black holes connected by ordinary strings. If the sides remain perfectly straight, they never self-intersect and the triangle undergoes chaotic oscillatory movement. It also shrinks by emitting gravitational waves. However, if the stringy sides support large-amplitude waves, the strings from different sides will likely reconnect which would lead to the triangle destruction. .}
    \label{fig:triangle1}
\end{figure}

Another example of a simple net is a triangular configuration with three BHs connected by three "ordinary" strings; see Fig.~(\ref{fig:triangle1}).  If the strings are nearly straight, they will remain straight, assuming that the BH motion is non-relativistic and that the BHs accelerate each other on a timescale longer than the light-crossing time for the triangle's side.  The strings will then never self-intersect.  However, if waves on strings are amplified by multiple reflections off BHs \cite{necklace}, the strings may reconnect and the triangle may be destroyed.

\begin{figure}
\includegraphics[width=\columnwidth]{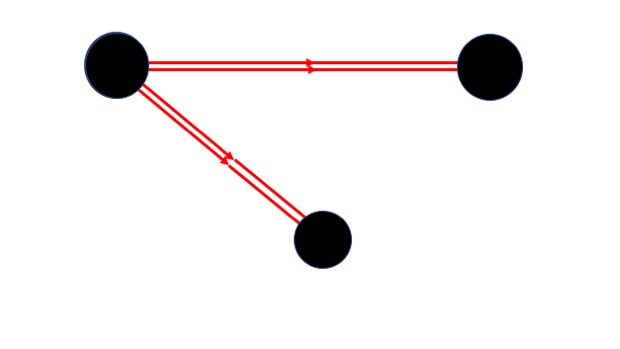}
    \caption{A stable triangle: three black holes with charges $4$, $-2$ and $-2$ are connected by $2$ pairs strings. The connections are topologically protected and are indestructible by the internal dynamics of the triple. The triple shrinks by emitting gravitational waves.}
    \label{fig:tripleprotected}
\end{figure}

A triplet net can also be formed for necklace-type strings, with three BHs having magnetic charges 2, 2 and -4. In this triple the BH with charge $4$ is connected to two black holes of charge $-2$ by two pairs of strings; see Fig. \ref{fig:tripleprotected}. This configuration is topologically protected, in a sense that the internal dynamics of the triple will not change its configuration even if the tension waves are excited on the strings stretched between the black holes.

In the maximal scenario, the rate of loop formation remains strongly suppressed even at $t>t_d$, but gravitational waves are still generated by the vibrating BH-string network.  The gravitational wave spectrum differs substantially between different scenarios.  Here we shall only consider some aspects of gravitational radiation by small string-BH nets.

\section{Gravitational radiation from nets}

Let's consider a small net of black holes of mass $M$ connected by strings with tension $\mu$.  A special case is a triplet of black holes, connected by straight strings. So long as the motion is non-relativistic, the strings segments remain straight, and since the sides of a triangle never intersect each other, the triangle can remain stable for many dynamical timescales.
Let $R$ be the characteristic size of the net. We shall work in the following approximations:

1.	$M\gg\mu R$. This means that BHs have more mass than strings and they move non-relativistically. This condition is satisfied at $t_d$ { for $G\mu\ll\lambda$ in the minimal scenario.}

2.	$G\mu\gg(GM/R)^2$. This means that the string tension is stronger than gravitational interaction between BHs. Numerically, for $G\mu\sim 10^{-10}$, this implies that the BHs are further away from each other than about $10^5$ Schwarzschild radii. This condition is satisfied at $t_d$ and later times.

Let us estimate the power radiated in gravitational waves. The quadrupole moment of the net is $Q\sim MR^2$ and $\ddot{Q}\sim Mv^2\sim \mu R$. The characteristic angular frequency of oscillations is 
\begin{equation}
\omega\sim v/R\sim\sqrt{\mu\over RM}=\sqrt{G\mu\over GM\times R}.
\end{equation}
The characteristic strain of   a gravitational wave at distance $r$ from the cluster is
\begin{equation}
h\sim G\ddot{Q}/r\sim G\mu R/r.
\end{equation}
One can find the energy loss rate from the emission of gravitational waves:
\begin{equation}
dE/dt\sim -h^2 r^2 \omega^2/G\sim (G\mu)^2 (\mu R)/(GM)\sim -(G\mu)^2(GM)^{-1}E
\end{equation}
Therefore, the energy of the net and its size decay approximately exponentially, on the timescale\footnote{A similar estimate was obtained in a different context in Ref.~\cite{JoseKen}.  Gravitational radiation from relativistically moving
masses connected by strings was studied in Refs.~\cite{MartinAV, Leblond}}
\begin{equation}
\tau_{\rm GW}\sim GM/(G\mu)^2\sim 10^{15}  \left({M\over M_{\odot}}\right)\left({10^{-10}\over G\mu}\right)^{-2}\hbox{sec}.
\label{GW}
\end{equation}.

Obviously, the expression above is a rough estimate and the precise
answer depends on the initial geometry of the system. Eventually the gravitational force between the BHs becomes greater than the string tension, either due to the overall shrinking of the net, or due to chaotic motion that would randomly bring two black holes to close separation. When BH gravitational interaction dominates, the the shrinking accelerates and  one obtains a merger, or a series of mergers.
Remarkably, the timescale in Eq.~(\ref{GW}) seemingly
implies that only BHs of certain mass scale are merging at the present 
epoch:

\begin{equation}
M\sim 100 M_{\odot}\Lambda 
\left({G\mu\over 10^{-10}}\right)^2 
\left({t_H\over 10^{17}\hbox{sec}}\right).
\end{equation}
where $t_H$ is the Hubble time and $\Lambda$ is the number of e-foldings by which the net needs to shrink before the BHs physically merge.
In other words, a certain mass scale is naturally selected by the age of the Universe and the string tension, and for realistic parameters this scale is within the mass range detectable by the current ground-based detectors and future space-based detectors. This argument, however, may be too simplistic: the chaotic dynamics of the net may produce bursts
of gravitational waves when the BHs pass close to each other, and this may cause a substantial variation in the merger timescale.
It is worth noting that the net possesses a characteristic oscillation frequency $f_0$ at which the force of gravitational attraction between the BHs is comparable to string tension:
\begin{equation}
f_0\sim {c^3\over GM} (G\mu)^{3/4}\sim 0.01\left({M_{\odot}\over M}\right)\left({G\mu\over 10^{-10}}\right)^{3/4}\hbox{Hz} 
\end{equation}
 For realistic values of the BH mass and string tension, this frequency could well lie within LISA or Pulsar Timing Array bands. The spectral shape of the stochastic gravitational-wave background that is produced by an ensemble of shrinking nets may have an observable feature at this frequency. 
 



A number of questions regarding small BH-string nets look interesting but are left unanswered: Does the chaotic nature of net dynamics significantly affect their lifetime and their GW spectrum?  In particular, how common are close encounters of BHs in a net, resulting in GW bursts or/and BH mergers?  What is the typical BH spin resulting from the torques due to the string tension?
How common are non-intersecting trajectories for loops of string attached to BHs?  Close to the points of loop attachment, some string length may escape through the BH horizon.  How important is this mechanism in depleting the loop's length over cosmological timescales?


\section{Constraints on persistent networks}

Suppose now that BHs remain connected to strings until the present time.  
The BH separation at present is then
\beq
d_0\sim \lambda^{-1/3}(t_f t_{eq})^{1/2} z_{eq} \sim 10^{22}\lambda_{12}^{-1/3}\left(\frac{M}{M_\odot}\right)^{1/2} {\rm cm},
\eeq
where $\lambda_{12}\equiv \lambda/10^{-12}$.  The energy density of strings relative to the background is
\beq
\frac{\rho_s}{\rho_0}\sim G\mu \frac{t_0^2}{d_0^2}\sim 10^{-2}\lambda^{2/3}G\mu\frac{t_0}{t_f}.
\eeq
Requiring that this is $< 1$, we obtain the constraint
\beq
G\mu < 
10^{-12} \lambda_{12}^{-2/3}\frac{M}{M_\odot}.
\label{rhoconstraint}
\eeq

For BHs to play a role in structure formation, we require that their string-induced velocities are sufficiently small:
\beq
v_0\sim (\mu d_0/M)^{1/2} \lesssim 100~{\rm km/s}.
\eeq
This gives the constraint
\beq
G\mu < 
10^{-19}\lambda_{12}^{1/3} \left(\frac{M}{M_\odot}\right)^{1/2}.
\label{vconstraint}
\eeq

For supermassive BHs with $M\sim 10^6 M_\odot$, the velocity constraint requires $G\mu\lesssim 10^{-16}$.  If this is satisfied, BHs would be captured by galaxies.  If the strings are superconducting \cite{Witten}, they could be observed as lines of radio emission attached to BHs \cite{Chudnovsky}.  {We note the intriguing possibility that the radio filament observed near the central BH in our Galaxy may be due to a cosmic string \cite{Morris}.}

For PBH dark matter with $M\sim 10^{-13} M_\odot$ and $\lambda\sim 10^{-15}$, the density and velocity constraints require $G\mu\lesssim 10^{-23}$ and $G\mu\lesssim 10^{-26}$, respectively.  Note that even if PBHs are captured in galaxies, they may avoid capture in much smaller halos, due to the pull of the strings.  PBHs attached to strings effectively act as warm dark matter; they may also help to explain the absence of overly dense cores at galactic centers (predicted by the standard $\Lambda$CDM). 

\section{conclusions}
In this work we introduced a new class of models in which cosmic strings and PBHs are both present. The interaction between the two is dramatic and results in the formation of an infinite string network with BHs at the nodes.  In the course of the following evolution, strings cross and reconnect, and some fraction of BHs is detached from the network.  
We have discussed two limiting scenarios, providing upper and lower bounds on the density of BHs that remain in the infinite network.  A more definitive analysis of network evolution will require numerical simulations.


One of the main observational signatures of cosmic strings is the gravitational wave (GW) background produced by oscillating string loops.  In the presence of PBHs, the loop production by the network is significantly modified, resulting in a modified GW spectrum.  In addition, the network provides another source of GW -- small oscillating nets of black holes and strings.  The nets lose their energy by gravitational radiation and shrink exponentially 
on a timescale of $ \tau_{GW}\sim t_H (M/100M_{\odot})(10^{-10}/G\mu)^{-2}$, where $t_H$ is the current age of the Universe. 
If such objects exist and are detected by LIGO, Pulsar Timing Arrays, or LISA, this would give most valuable insights for particle physics and astrophysics alike.  To further investigate observational signatures of the model, the properties of nets with black holes need to be explored in greater detail.  



\begin{acknowledgements}

We are grateful to Ken Olum, Jose Blanco-Pillado, and Elena Murchikova for useful discussions and comments on the manuscript.
The work of A.V. was supported by the National Science Foundation.

\end{acknowledgements}


\begin{thebibliography}{99}

\bibitem{Carr:2009jm} 
  B.~J.~Carr, K.~Kohri, Y.~Sendouda and J.~Yokoyama,
  ``New cosmological constraints on primordial black holes,''
  Phys.\ Rev.\ D {\bf 81}, 104019 (2010)
    [arXiv:0912.5297 [astro-ph.CO]].

\bibitem{Sasaki:2018dmp} 
  M.~Sasaki, T.~Suyama, T.~Tanaka and S.~Yokoyama,
  ``Primordial black holes?perspectives in gravitational wave astronomy,''
  Class.\ Quant.\ Grav.\  {\bf 35}, no. 6, 063001 (2018)
    [arXiv:1801.05235 [astro-ph.CO]].

\bibitem{book} 
  A.~Vilenkin and E.~P.~S.~Shellard,
  {\it Cosmic Strings and Other Topological Defects}
(Cambridge University Press, Cambridge, 2000).

\bibitem{Copeland} 
  E.~J.~Copeland and T.~W.~B.~Kibble,
  ``Cosmic Strings and Superstrings,''
  Proc.\ Roy.\ Soc.\ Lond.\ A {\bf 466}, 623 (2010)
    [arXiv:0911.1345 [hep-th]].

\bibitem{Zeldovich}
Ya.~B.~Zeldovich and I.~D.~Novikov, 
Sov.\ Astron. {\bf 10}, 602 (1967).

\bibitem{CarrHawking} 
  B.~J.~Carr and S.~W.~Hawking,
  Mon.\ Not.\ Roy.\ Astron.\ Soc.\  {\bf 168}, 399 (1974).

\bibitem{GarciaBellido:1996qt} J.~García-Bellido, A.~D.~Linde
and D.~Wands, ``Density perturbations and black hole formation in
hybrid inflation,'' Phys.\ Rev.\ D \textbf{54}, 6040 (1996) {[}astro-ph/9605094{]}.

\bibitem{Yokoyama:1995ex} J.~Yokoyama, ``Formation of MACHO primordial
black holes in inflationary cosmology,'' Astron.\ Astrophys.\ \textbf{318},
673 (1997) {[}astro-ph/9509027{]}.

\bibitem{Frampton:2010sw} P.~H.~Frampton, M.~Kawasaki, F.~Takahashi
and T.~T.~Yanagida, ``Primordial Black Holes as All Dark Matter,''
JCAP \textbf{1004}, 023 (2010) {[}arXiv:1001.2308 {[}hep-ph{]}{]}.


\bibitem{Carr:2016drx} 
  B.~Carr, F.~Kuhnel and M.~Sandstad,
  ``Primordial Black Holes as Dark Matter,''
  Phys.\ Rev.\ D {\bf 94}, no. 8, 083504 (2016)
    [arXiv:1607.06077 [astro-ph.CO]].

\bibitem{Inomata:2017vxo} 
  K.~Inomata, M.~Kawasaki, K.~Mukaida and T.~T.~Yanagida,
  ``Double inflation as a single origin of primordial black holes for all dark matter and LIGO observations,''
  Phys.\ Rev.\ D {\bf 97}, no. 4, 043514 (2018)
   [arXiv:1711.06129 [astro-ph.CO]].

\bibitem{Sasaki:2016jop} 
  M.~Sasaki, T.~Suyama, T.~Tanaka and S.~Yokoyama,
  ``Primordial Black Hole Scenario for the Gravitational-Wave Event GW150914,''
  Phys.\ Rev.\ Lett.\  {\bf 117}, no. 6, 061101 (2016)
    [arXiv:1603.08338 [astro-ph.CO]].

\bibitem{GVZ} J.~Garriga, A.~Vilenkin and J.~Zhang, ``Black holes
and the multiverse,'' JCAP \textbf{1602}, no. 02, 064 (2016) {[}arXiv:1512.01819
{[}hep-th{]}{]}.

\bibitem{DV} 
  H.~Deng and A.~Vilenkin,
  ``Primordial black hole formation by vacuum bubbles,''
  JCAP {\bf 1712}, no. 12, 044 (2017)
    [arXiv:1710.02865 [gr-qc]].

\bibitem{Garriga:1992nm} J.~Garriga and A.~Vilenkin, ``Black holes
from nucleating strings,'' Phys.\ Rev.\ D \textbf{47}, 3265 (1993)
{[}hep-ph/9208212{]}.

\bibitem{Khlopov:2004sc} M.~Y.~Khlopov, S.~G.~Rubin and A.~S.~Sakharov,
``Primordial structure of massive black hole clusters,'' Astropart.\ Phys.\ \textbf{23},
265 (2005) {[}astro-ph/0401532{]}.

\bibitem{Yoo} 
  N.~Tanahashi and C.~M.~Yoo,
  ``Spherical Domain Wall Collapse in a Dust Universe,''
  Class.\ Quant.\ Grav.\  {\bf 32}, no. 15, 155003 (2015)
    [arXiv:1411.7479 [gr-qc]].

\bibitem{DGV} H.~Deng, J.~Garriga and A.~Vilenkin, ``Primordial
black hole and wormhole formation by domain walls,'' JCAP \textbf{1704},
no. 04, 050 (2017) {[}arXiv:1612.03753 {[}gr-qc{]}{]}.

\bibitem{Hawking:1987bn} S.~W.~Hawking, ``Black Holes From Cosmic
Strings,'' Phys.\ Lett.\ B \textbf{231}, 237 (1989).

\bibitem{Polnarev:1988dh} A.~Polnarev and R.~Zembowicz, ``Formation
of Primordial Black Holes by Cosmic Strings,'' Phys.\ Rev.\ D \textbf{43},
1106 (1991).


\bibitem{Carr:2017edp} 
  B.~Carr, T.~Tenkanen and V.~Vaskonen,
  ``Primordial black holes from inflaton and spectator field perturbations in a matter-dominated era,''
  Phys.\ Rev.\ D {\bf 96}, no. 6, 063507 (2017)
    [arXiv:1706.03746 [astro-ph.CO]].

\bibitem{Bennett} 
  D.~P.~Bennett and F.~R.~Bouchet,
  ``Cosmic string evolution,''
  Phys.\ Rev.\ Lett.\  {\bf 63}, 2776 (1989).

\bibitem{Allen} 
  B.~Allen and E.~P.~S.~Shellard,
  ``Cosmic string evolution: a numerical simulation,''
  Phys.\ Rev.\ Lett.\  {\bf 64}, 119 (1990).

\bibitem{BlancoPillado:2011dq} 
  J.~J.~Blanco-Pillado, K.~D.~Olum and B.~Shlaer,
  ``Large parallel cosmic string simulations: New results on loop production,''
  Phys.\ Rev.\ D {\bf 83}, 083514 (2011)
    [arXiv:1101.5173 [astro-ph.CO]].

\bibitem{Blanco-Pillado:2017rnf} 
  J.~J.~Blanco-Pillado, K.~D.~Olum and X.~Siemens,
  ``New limits on cosmic strings from gravitational wave observation,''
  Phys.\ Lett.\ B {\bf 778}, 392 (2018)
    [arXiv:1709.02434 [astro-ph.CO]].

\bibitem{BOS} 
  J.~J.~Blanco-Pillado, K.~D.~Olum and B.~Shlaer,
  ``Cosmic string loop shapes,''
  Phys.\ Rev.\ D {\bf 92}, no. 6, 063528 (2015)
    [arXiv:1508.02693 [astro-ph.CO]].

\bibitem{Kibble:2015twa} 
  T.~W.~B.~Kibble and T.~Vachaspati,
  ``Monopoles on strings,''
  J.\ Phys.\ G {\bf 42}, no. 9, 094002 (2015)
    [arXiv:1506.02022 [astro-ph.CO]].

\bibitem{necklace} 
  X.~Siemens, X.~Martin and K.~D.~Olum,
  ``Dynamics of cosmic necklaces,''
  Nucl.\ Phys.\ B {\bf 595}, 402 (2001)
    [astro-ph/0005411].

\bibitem{BPO} 
  J.~J.~Blanco-Pillado and K.~D.~Olum,
  ``Monopole annihilation in cosmic necklaces,''
  JCAP {\bf 1005}, 014 (2010)
    [arXiv:0707.3460 [astro-ph]].

\bibitem{JoseKen}
J.~J.~Blanco-Pillado and K.~D.~Olum,
  ``Monopole - anti-monopole bound states as a source of ultrahigh-energy cosmic rays,''
  Phys.\ Rev.\ D {\bf 60}, 083001 (1999)
    [astro-ph/9904315].

\bibitem{MartinAV} 
  X.~Martin and A.~Vilenkin,
  ``Gravitational radiation from monopoles connected by strings,''
  Phys.\ Rev.\ D {\bf 55}, 6054 (1997)
    [gr-qc/9612008].

\bibitem{Leblond} 
  L.~Leblond, B.~Shlaer and X.~Siemens,
  ``Gravitational Waves from Broken Cosmic Strings: The Bursts and the Beads,''
  Phys.\ Rev.\ D {\bf 79}, 123519 (2009)
    [arXiv:0903.4686 [astro-ph.CO]].

\bibitem{Witten} 
  E.~Witten,
  ``Superconducting Strings,''
  Nucl.\ Phys.\ B {\bf 249}, 557 (1985).

\bibitem{Chudnovsky} 
  E.~M.~Chudnovsky, G.~B.~Field, D.~N.~Spergel and A.~Vilenkin,
  ``Superconducting Cosmic Strings,''
  Phys.\ Rev.\ D {\bf 34}, 944 (1986).

\bibitem{Morris}
M.~R.~Morris, J.-H.~Zhao and W.~M.~Goss,
"A nonthermal radio filament connected to the galactic black hole?,"
[arXiv:1711.04190 [astro-ph]]


\end{thebibliography}
\end{document}